\documentclass[submission,copyright,creativecommons]{eptcs}
\providecommand{\event}{ACL2 2023} 
\usepackage{underscore}           

\usepackage{graphicx}
\usepackage{xcolor} 
\usepackage{amsmath}
\usepackage{amsthm}
\usepackage{amsfonts}
\usepackage{amssymb}

\usepackage{listings}
\definecolor{commentsColor}{rgb}{0.497495, 0.497587, 0.497464}
\definecolor{keywordsColor}{rgb}{0.000000, 0.000000, 0.635294} 
\definecolor{stringColor}{rgb}{0.558215, 0.000000, 0.135316}
\title{Using Counterexample Generation and Theory Exploration to Suggest Missing Hypotheses}
\author{Ruben Gamboa
\institute{University of Wyoming \\ \& Kestrel Institute\thanks{This work was
supported by a grant from the Defense Advanced Research Projects Agency (DARPA) Proof Engineering, Adaptation, Repair, and Learning for Software (PEARLS) Artificial Intelligence Exploration (AIE) Opportunity} \\ Laramie, Wyoming}
\email{ruben@uwyo.edu}
\and
Panagiotis Manolios
\institute{Northeastern University \\ Boston, Massachusetts}
\email{p.manolios@northeastern.edu}
\and
Eric Smith
\institute{Kestrel Institute\footnotemark[1]  \\ Palo Alto, California}
\email{eric.smith@kestrel.edu}
\and
Kyle Thompson
\institute{University of California San Diego  \\ \& Kestrel Institute\footnotemark[1]  \\ San Diego, California}
\email{r7thompson@ucsd.edu}
}
\def\titlerunning{Suggesting Missing Hypotheses}
\def\authorrunning{Gamboa, Manolios, Smith, Thompson}
\begin{document}
\maketitle

\begin{abstract}
Newcomers to ACL2 are sometimes surprised that ACL2 rejects formulas that they believe should be theorems, such as \texttt{(reverse (reverse x)) = x}. Experienced ACL2 users will recognize that the theorem only holds for intended values of \texttt{x}, and given ACL2's total logic, there are many counterexamples for which this formula is simply not true. Counterexample generation (cgen) is a technique that helps by giving the user a number of counterexamples (and also witnesses) to the formula, e.g., letting the user know that the intended theorem is false when \texttt{x} is equal to 10. In this paper we describe a tool called DrLA that goes further by suggesting additional hypotheses that will make the theorem true. In this case, for example, DrLA may suggest that \texttt{x} needs to be either a \texttt{true-list} or a \texttt{string}. The suggestions are discovered using the ideas of theory exploration and subsumption from automated theorem proving.
\end{abstract}

\section{Introduction}
\label{sec:intro}

Over the past year, the Kestrel PEARLS team has been working to implement
ideas to use machine learning tools to improve the experience of users in 
the construction and repair of proofs. Most of these ideas revolve around
The Method~\cite{acl2::the-method}. In particular, we built an ``advice''
tool that can read an ACL2 checkpoint from a failed proof attempt, and 
suggest a number of routes the user may take to resolve the issue. The
advice is created from a variety of models trained using machine learning
techniques, as well as some heuristics that would be familiar to ACL2
users. The models are trained by taking data from the ACL2 Community
Books, deliberating breaking the theorems in those books, and submitting the
broken theorem to ACL2. The model is trained to recognize the checkpoint 
that ACL2 discovers when trying to prove the broken theorem, and then suggest
the fix that corresponds to the way the theorem was originally broken. For
example, one way to break a theorem is to remove an \texttt{include-book}, 
so the proposed fix is to include that particular library book. 

There are, of course, various ways to break theorems, thus various different
solutions that the advice tool may suggest. Besides removing library books,
we may remove hints, or remove hypotheses from the theorem itself. Two 
lessons we learned
while doing this are that (1) sometimes the advice tool could
produce valuable advice using only hardcoded suggestions instead of
full-blown machine learning (e.g., ``try enabling all definitions''), and
(2) advising the user to add a hypothesis is inherently riskier than 
suggesting including a book or adding a hint. The second point is simply
unavoidable, since adding a hypothesis \emph{changes} the logical meaning
of the intended theorem. For instance, suggesting the hypothesis \texttt{NIL}
will result in a successful proof attempt of a useless theorem. But a less
blatant problem would be suggesting the new hypothesis \texttt{x <= 0} in a 
theorem that already has the hypothesis \texttt{x >= 0}; the new theorem, 
which applies only to the case \texttt{x = 0} may be easier to prove, but it 
is also much less useful.

In light of this, and further considering the first point above, it is 
natural to ask whether there are simple strategies---i.e., not based on 
machine learning or artificial intelligence in general---that can be 
used effectively to suggest missing hypotheses. This paper introduces
DrLA, a tool that does precisely this. Rather than use checkpoints from a
failed proof attempt, DrLA infers missing hypotheses by using the 
counterexample generation engine (cgen) originally developed in 
ACL2s~\cite{Pete:cgen}. Cgen provides the user with both counterexamples
and witnesses to the proposed theorem, and DrLA uses theory exploration
techniques~\cite{Johansson-2021} to suggest the missing hypotheses. A key
concept in theory exploration is to consider only terms that do not reduce
to terms that have already been seen. In the context of generating
hypotheses, this deals effectively with the problem of suggesting hypotheses 
that trivialize the original theorem.

The rest of the paper is organized as follows. Sect.~\ref{sec:background}
describes the necessary background from theory exploration to then introduce
the key ideas behind hypothesis generation with DrLA. This is followed in
Sect.~\ref{sec:assessment} with a discussion of the effectiveness
of DrLA (and similar tools). Then Sect.~\ref{sec:implementation} discusses
some details of the DrLA implementation. Finally, Sect.~\ref{sec:conclusion}
provides some concluding remarks and suggests avenues for the future
evolution of DrLA and other tools to ease proof development with ACL2.

\section{Background and the Key Idea}
\label{sec:background}

Theory exploration is a technique for discovering likely properties of
programs or lemmas of a mathematical theory. For example, once the function
\texttt{append} over lists is defined, theory exploration may discover that
append is, in fact, associative. This is done by combining two tools: a
formula generator and a property checker (also known as a counterexample
generator). 

The formula generator creates formulas, i.e., possible theorems, from a 
given set of function and constant symbols. In the case of list functions, 
it may start with
\texttt{consp},
\texttt{nil},
\texttt{cons},
\texttt{car},
\texttt{cdr},
\texttt{append}, and
\texttt{equal}.
Using this vocabulary, the theory exploration tool may create some
familiar theorems such as
\begin{itemize}
\item \texttt{(equal (car (cons x1 x2)) x1)}
\item \texttt{(equal (append (append x1 x2) x3) (append x1 (append x2 x3)))}
\end{itemize}
as well as reasonable-looking formulas that are not theorems, e.g.,
\begin{itemize}
\item \texttt{(equal (car (cons x1 x2)) x2)}
\item \texttt{(equal (append x1 x2) (append x2 x1))}
\end{itemize}
and complete nonsense, such as
\begin{itemize}
\item \texttt{(consp (equal (car nil) (append x1 x2)))}
\item \texttt{(car (cons (cdr x1) (equal x2 nil)))}
\end{itemize}

Theory exploration then considers each of these formulas in turn, and 
determines which of them are likely to be true. This is where the property
checker comes in, by methodically searching for counterexamples to each
formula. 
E.g., the first conjecture is \texttt{(equal (car (cons x1 x2)) x1)}, and
it has the variables \texttt{x1} and \texttt{x2}, so the property checker
will consider thousands of random or strategically chosen values for them, 
such as \texttt{x1=3}, \texttt{x2='(1 2)} or \texttt{x1='(a~.~16)}, 
\texttt{x2='bgs}. In all cases, the formula ends up
being true, so theory exploration will suggest this formula as a 
\emph{likely} lemma. Theorem proving can then be used to confirm that it is 
an \emph{actual} lemma, and some theory exploration systems do this.

In the case of spurious theorems, theory exploration can often find an
assignment that demonstrates the formula cannot be true. E.g., the binding
\texttt{x1=3}, \texttt{x2='(1 2)} from above suffices to show that the
formula \texttt{(equal (car (cons x1 x2)) x2)} cannot be true, so it
would never be suggested as a likely lemma. The same is true of the
nonsensical formulas, though care must be taken with respect to runtime
errors, since these expressions may violate guards freely.

So the result of theory exploration is a list of theorems, or at least
likely conjectures. 
The goal is to produce enough formulas that the tool can find a sufficient
number of useful theorems. 
Obviously there is a 
delicate balance involving the formula generator. Ideally, it should
generate as many formulas as possible, so that useful lemmas can be
discovered. But the process of generating random formulas grows
exponentially with their length,
so limits are unavoidable, and efficient strategies are used to prune
the space of candidate formulas so that barren areas of the search space
are not explored.
In practice, this means that the theorems discovered are usually small 
syntactically, e.g., limited in terms of depth. 

For our purposes, we are interested in discovering not likely lemmas, but
likely hypotheses that may be missing from a theorem. The overall strategy
remains the same: A term generator will produce candidate hypotheses, and
a property checker can determine if each possible hypothesis is likely to make the theorem provable. But a key idea is that we can leverage the work of the property
checker since the hypotheses are always in the context of a surrounding
formula, as opposed to theory exploration where the generated formulas are
all at the top level. E.g., consider the motivating example
\texttt{(equal (reverse (reverse x)) x)}. It is a reasonable heuristic to expect 
that any missing hypothesis will feature only the variable \texttt{x}, so
we can generate values for \texttt{x} ahead of time and test all candidate
hypotheses with the same set of bindings.

In fact, we can do a bit better than that. Cgen, the counterexample
generator developed as part of ACL2s, is a sophisticated tool that will
find both counterexamples and witnesses to an ACL2 formula. For our 
motivating example, cgen will identify the following counterexamples
\begin{itemize}
\item \texttt{((X '((T . 1) NIL . \#$\backslash$A)))}
\item \texttt{((X '(-25 . 0)))}
\item \texttt{((X '(53 . 252)))}
\end{itemize}
and the following witnesses
\begin{itemize}
\item \texttt{((X '((T T) (\#$\backslash$A 1))))}
\item \texttt{((X NIL))}
\item \texttt{((X '(-1)))}
\end{itemize}
Readers experienced with ACL2 will immediately recognize that all of the
witnesses are true lists, whereas none of the counterexamples are---which
immediately suggests \texttt{(true-listp x)} as the missing hypothesis.

DrLA proceeds in a similar manner. The basic idea is to find an expression
that is false for all of the counterexamples and true for all 
witnesses\footnote{We will see later that it is not necessary, or even
desirable, for the property to hold for all witnesses, but this is a good 
first approximation.}. This evokes the machine learning idea of finding
a ``hyperplane'' that separates the positive and negative examples in a
training data set. This is straightforward to do with a general property
checker, as in theory exploration.

The more interesting component is the term generator. What should be the
language (i.e., function symbols) that determines the possible terms that
lead to possible hypotheses? It is often the case that the missing
hypothesis is a type hypothesis. ACL2 is an untyped language, but many
functions are written with specific types in mind. This is certainly the
case where \texttt{reverse} is concerned, since the obvious programmer
intent is for \texttt{reverse} to work with lists, more specifically
true-lists. Theorems about these functions usually require some typing
hypothesis to make explicit the intended use of the function, and such
hypotheses are easy to miss.

The tau system is an important component of ACL2 that helps the theorem 
prover benefit from the implicit notions of type assumed by 
programmers~\cite{acl2::tau}. Tau is designed ``to be a lightweight, fast, 
and helpful decision procedure for an elementary subset of the logic focused 
on monadic predicates and function signatures''~\cite{acl2::tau}. So a
reasonable language for the term generator is the set of types (``monadic
predicates'') in use by the tau system. This set begins with a hard-coded
list of primitive predicates, including \texttt{consp}, \texttt{natp}, and so 
on. The set is enhanced when new definitions are encountered, e.g., 
\texttt{primep}, but only monadic predicates are considered. To account for
the fact that sometimes the result of comparisons is a useful notion of type,
tau hardcodes some common comparisons as primitive type predicates, e.g.,
\texttt{0 <= x}, \texttt{1 < x}, and ``\texttt{x} is a non-NIL true-listp.''
DrLA adopts the primitive types recognized by the tau system, but unlike the
tau system, it does not currently expand this dictionary as new functions 
are introduced. Rather, it uses the the predicates mentioned in the theorem
and used in the definitions of functions present in the theorem. For example,
when considering \texttt{(equal (reverse (reverse x)) x)}, it will add the
function \texttt{reverse} to the list of types, as well as functions used
in its definition such as \texttt{revappend}. This illustrates another
departure from the tau system, in that \texttt{revappend} is a binary
function, so it would be ignored by the tau system.

DrLA will take these selected symbols and generate terms by nesting
syntactically valid function invocations. These terms are the expression
trees that can be generated using these function symbols up to a
maximum depth. The leaves of the terms correspond to either variable symbols,
which must occur in the original theorem, or one of a list of predefined
constant symbols, e.g., 0 or NIL. 
Thus, DrLA will explore terms including the following:
\begin{itemize}
\item \texttt{(posp x)}
\item \texttt{(consp x)}
\item \texttt{(reverse x)}
\item \texttt{(revappend x 0)}
\item \texttt{(equal (reverse x) x)}
\end{itemize}
DrLA will also consider boolean combinations of these terms, up to a
maximum depth. For example, DrLA may consider the hypothesis
\texttt{(or (posp x) (consp x))}. DrLA allows the user to control all of
the depth parameters: how many levels of function definitions to explore for
new function names, how deeply to nest boolean expressions, and how deeply
to nest non-boolean terms such as \texttt{(reverse (reverse x))}.

We will have more to say about the implementation of DrLA in
Sect.~\ref{sec:implementation}, but it is important to note now that
DrLA attempts to avoid unnecessary computation when exploring the search
space. For instance, DrLA is aware that the boolean predicates \texttt{and}
and \texttt{or} are commutative, so it considers only one of the terms
\texttt{(and P Q)} or \texttt{(and Q P)}. Moreover, it avoids nesting the
primitive predicates, so that it will consider
\texttt{(posp (reverse x))} but never
\texttt{(posp (consp x))}.

As described thus far, DrLA will respond to the motivating example with
an excess of possible hypotheses. Included in this list is the expected
\texttt{(true-listp x)}. But DrLA also finds other possible suggestions,
such as
\begin{itemize}
\item \texttt{(equal x 'nil)}
\item \texttt{(equal (revappend x x) 'nil)}
\item \texttt{(true-listp (revappend x x))}
\item \texttt{(and (consp x) (true-listp x))}
\item \texttt{(and (true-listp x) (equal (reverse x) 'nil))}
\end{itemize}
Actually, DrLA finds dozens of similar, unhelpful suggestions.

Thus, the final component of DrLA is a filter that reduces the number of
suggestions in a manner reminiscent of subsumption. Specifically, suppose
that DrLA has two suggestions $P$ and $Q$ such that $P\rightarrow Q$ but
$Q\not\rightarrow P$. For example, $P$ may be \texttt{(natp x)} while
$Q$ is \texttt{(integerp x)}. We say in this case that $Q$ is more general 
(logically weaker)
than $P$, and we prefer to suggest $Q$. With this heuristic, DrLA will
suggest \texttt{(true-listp x)} but not \texttt{(equal x 'nil)} since 
\texttt{(true-listp x)} is the more general term. This heuristic eliminates
many of the useless suggestions above, but not all. For instance,
\texttt{(true-listp (revappend x x))} is logically equivalent to
\texttt{(true-listp x)}, so neither is more general than the other.
In these cases, a complexity heuristic comes into play---if $P$ and $Q$ are
logically equivalent and $Q$ is syntactically simpler than $P$, DrLA will
suggest $Q$ but not $P$. Our notion of syntactic complexity is simple and
partial, so it is possible that DrLA will find two suggestions that are
logically equivalent and just as simple syntactically. In these cases, 
DrLA will offer both suggestions, letting the user pick which one to use as
the ``better'' hypothesis.

As mentioned previously, this notion of ``more general'' is similar to the
notion of subsumption in resolution theorem proving. In the implementation
of DrLA, it is used in much the same way. First, a suggestion is ignored
if it is subsumed by a prior suggestion (c.f.~forward subsumption.) 
Then, when a suggestion is added to the list of suggestions, prior 
suggestions subsumed by the new one are discarded (c.f.~backward 
subsumption.)

We note briefly that these heuristics also have a useful side-effect. In
Sect.~\ref{sec:intro}, we warned about the dangers of suggesting vacuous
hypotheses, such as \texttt{nil}. In fact, DrLA would never make such a
suggestion, since such hypotheses would never be true of all the witnesses.
But DrLA may suggest overly constrained hypotheses, such as
\texttt{(equal x 'nil)}. However, the heuristics described will rule out
that hypothesis in favor of the more general \texttt{(true-listp x)}.  

With these heuristics, DrLA suggests a single hypothesis to the user, but it
is \emph{not} the expected \texttt{(true-listp x)}. Instead, DrLA suggests
that the correct theorem is
\begin{verbatim}
    (implies (or (stringp x) 
                 (true-listp x))
             (equal (reverse (reverse x)) x))
\end{verbatim}
We believe that most users of ACL2 think of \texttt{reverse} as a function
that operates on lists, but in fact the programmers of this function
allowed for both lists and \emph{strings}. The heuristics of DrLA select
this hypothesis since it is more general than the expected 
\texttt{(true-listp x)}.

\section{Assessment}
\label{sec:assessment}

In this section we describe our efforts to assess the performance of DrLA.
The approach is to give DrLA versions of theorems from the Community Books,
after removing one or more of the hypotheses in the theorem. This is the
same idea used to assess the more general advice tool that we built as part
of the PEARLS project, but with an important difference. The other suggested
fixes address the proof itself, e.g., by suggesting a hint. DrLA, on the other
hand, suggests a  modification to the theorem, e.g., a new hypothesis to add.
DrLA is never invoked unless cgen demonstrates that the theorem is false, i.e.,
by finding some counterexamples.

What this means for assessment is that it is insufficient simply to check if
the the suggestion results in a successful proof attempt. What is really
important is to determine whether the suggestion matches the original intent
of the ACL2 user, who presumably forgot to list one of more key hypotheses.
Simply comparing the suggested hypothesis with the original one, i.e., the one
that was removed earlier, is also insufficient, because there may be more than 
one way to express the necessary constraint.

So we chose to do much of the assessment manually, by running DrLA on examples
from the Community Books and determining whether DrLA's output is effective.
For example, the book \url{std/lists/append.lisp} is part of the lists library
in ACL2, and it starts with the theorem
\begin{verbatim}
    (implies (consp x)
             (< 0 (len x)))
\end{verbatim}
Removing the hypothesis leaves just \texttt{(< 0 (len x))}, and when this is
submitted to DrLA, it provides a single suggestion: \texttt{(consp x)}. 
Moreover, DrLA reports that after adding this hypothesis to the theorem, ACL2
is indeed able to find a proof. In this specific example, DrLA's performance
is an unqualified success.

The next theorem in the file is
\begin{verbatim}
    (implies (not (consp x))
             (equal (append x y)
                    y))
\end{verbatim}
When DrLA is prompted with the conclusion, it suggests \texttt{(atom x)}
as the missing hypothesis. Note that \texttt{atom} is defined in ACL2 as
not \texttt{consp}, so this is 100\% consistent with the original
theorem, although not identical. This is why we think a manual assessment
process is necessary.

A more interesting theorem is
\begin{verbatim}
    (iff (append x y)
         (or (consp x)
             (consp y)))
\end{verbatim}
DrLA does not support \texttt{iff}, but we can use this by breaking the
theorem up into two implications. When presented with \texttt{(append x y)},
DrLA accurately suggests \texttt{(or (consp x) (consp y))}, but it also
makes a number of other suggestions which, while valid, are less useful.
For instance, it suggests
\texttt{(or (acl2-numberp y) (consp x))}.
Notice that neither of these suggestions subsumes the other, and neither is
syntactically simpler than the other, so DrLA suggests 
both\footnote{Actually, DrLA gives six total suggestions in this case.} and 
lets the user decide which is the best alternative.

In the other direction, DrLA is completely ineffective. There are many 
witnesses and counterexamples for \texttt{(or (consp x) (consp y))}, but
DrLA fails to find a single suggestion. The reason is that cgen finds many
counterexamples to this formula, but it actually does not find any 
witnesses. In any case, the conclusion is a disjoint of simple types, so
DrLA would be hard-pressed to find a \emph{simpler} formula to suggest.
It would certainly not even explore terms with the function symbol
\texttt{append}.

The situation is somewhat better in the case of the following theorem:
\begin{verbatim}
    (implies (and (not (index-of k x))
                  (index-of k y))
             (equal (index-of k (append x y))
                    (+ (len x) (index-of k y))))
\end{verbatim}
When DrLA is presented with just the conclusion of this theorem, it 
discovers the missing hypotheses
\texttt{(and (not (index-of k x)) (index-of k y))}.
This reason this succeeds, is precisely that the key predicate, 
\texttt{index-of},
appears in the conclusion of the theorem, so DrLA knows to generate 
possible hypotheses that involve \texttt{index-of}.

In contrast, consider the following
theorem, also from the standard lists library:
\begin{verbatim}
    (implies (member k x)
             (equal (nth (index-of k x) x)
                    k))	
\end{verbatim}
DrLA will consider terms that feature the primitive typing predicates, as 
well as any functions used in the theorem and the definition of functions
that appear in the theorem, e.g., \texttt{nth} and \texttt{index-of}. However,
\texttt{member} does not appear in the theorem or any of those definitions, so 
DrLA will fail to suggest the hypothesis of the theorem. These situations, 
where the missing hypothesis requires a predicate that does not otherwise 
appear in the theorem, are common.

To address this, DrLA allows the user to provide a list of additional predicate
symbols that it should also consider. If \texttt{member} is provided in this
way, DrLA may suggest \texttt{(member k x)} as a possible missing hypothesis.
Unfortunately, DrLA's subsumption heuristic leads DrLA astray in this scenario.
The problem is that any term that looks like \texttt{(or (member k x) P)}
subsumes \texttt{(member k x)} regardless of what \texttt{P} is. Thus, DrLA
suggests terms like the following instead of the correct
\texttt{(member k x)}:
\begin{itemize}
\item \texttt{(or (member k x) (and (rationalp x) (< x 0)))}
\item \texttt{(or (complex-rationalp x) (member k x))}
\end{itemize}
But in fact, these weaker hypotheses are not sufficient to prove the theorem;
when \texttt{x} is 1/2, for example, 
\texttt{(nth (index-of k x) x)} is equal to \texttt{nil}, not necessarily
equal to \texttt{k}. But the reason this shows up as a candidate missing
hypotheses is that the candidates are chosen empirically by examining the
counterexamples and witnesses---not by invoking the theorem prover directly.
What DrLA lacks here is a large enough sample of witnesses and 
counterexamples, e.g., at least one counterexample where \texttt{x} is a
rational or complex rational. So DrLA provides the user with 11 total
suggestions, though it does inform the user that only two of them are 
sufficient to prove the original theorem, and those two end up being vacuous,
i.e., with \texttt{x=nil}.

As these examples make clear, it is vital to the success of DrLA to start with
a suitable set of function symbols so the forest of 	possible hypotheses is rich
enough to contain useful candidates. Allowing the user to provide some 
candidates helps DrLA to succeed in some difficult cases, but it also feels
inappropriate: If the user knows that \texttt{member} is a suitable hypothesis,
surely she can simply provide the necessary hypothesis without using DrLA at 
all. This is an open issue that we will return to.

\section{Implementation Notes}
\label{sec:implementation}

In this section, we describe some of the implementation decisions and 
tradeoffs that we encountered implementing DrLA. As previously mentioned,
the first step is to invoke cgen using the function \texttt{prove/cgen} to
generate a list of witnesses and counterexamples, and we ask cgen for 50
of each. If cgen cannot find any witnesses or any counterexamples, DrLA 
reports an error and does not attempt to find missing hypotheses.

Otherwise, DrLA proceeds by generating the expressions that may become
candidate hypotheses. Although there is no such distinction in ACL2, DrLA
considers Boolean patterns, predicates, and terms separately. Boolean patterns
are partial expressions such as
\begin{itemize}
\item \texttt{(or NIL NIL)}
\item \texttt{(and (or NIL NIL) NIL)}
\item \texttt{(and (not NIL) NIL)}
\item \texttt{NIL}
\end{itemize}
The \texttt{NIL}s are placeholders where an arbitrary predicate can be
placed. These are just slots; there are no restrictions on the way predicates 
may appear, and in particular there is no requirement that the same predicate 
be used to replace multiple \texttt{NIL}s. Note: The significance of the
last template, a single \texttt{NIL}, will become clear shortly.

DrLA then collects the primitive type predicates (as in the tau system),
essential comparators (such as \texttt{equal} and \texttt{<<}), and any
extra predicates suggested by the user (e.g., \texttt{member} above). Then
it generates templates, similar to the Boolean patterns, using these names.
For example, DrLA may generate terms such as
\begin{itemize}
\item \texttt{(equal NIL NIL)}	
\item \texttt{(integerp NIL)}	
\item \texttt{(member NIL NIL)}	
\end{itemize}
Unlike the case with the Boolean expressions, DrLA does not nest these 
templates, since they are intended to be Booleans that operate on terms.
(Recall that DrLA makes a distinction between Boolean patterns, predicates,
and general terms.) The intent is to avoid considering unpromising expressions
such as \texttt{(integerp (memberp X (equal Y Z)))}.

DrLA then combines these two template lists into a single list of templates, 
which may include such entries as
\begin{itemize}
\item \texttt{(equal NIL NIL)}	
\item \texttt{(or (equal NIL NIL) (integerp NIL))}
\end{itemize}
Note that each template is created by replacing one of the \texttt{NIL}s in
the Boolean templates by one of the predicate templates (and now the reason
for the single \texttt{NIL} Boolean template should be clear.) There is an
exponential explosion here, so care is taken to use only tail-recursive
functions to avoid running out of stack space. Using DrLA's default values,
the total number of templates formed at this stage is 2,380---but that number
can go up, e.g., if the user specifies a larger depth for boolean expressions.

Next, DrLA generates the term templates, which are similar to the above, but
based on the functions that are present in the theorem or that appear
in the definitions of those functions, up to a certain depth limit. For
example, for the theorem 
\begin{verbatim}
    (equal (index-of k (append x y))
           (+ (len x) (index-of k y)))
\end{verbatim}
DrLA will construct terms from the functions
\texttt{APPEND},
\texttt{INDEX-OF},
\texttt{+},
\texttt{CDR},
\texttt{LEN}, and
\texttt{BINARY-+}.
Thus, it will generate the following 13 templates:
\begin{itemize}
\item \texttt{(+ NIL NIL)}
\item \texttt{(append NIL NIL)}
\item \texttt{(binary-+ NIL NIL)}
\item \texttt{(car NIL)}
\item \texttt{(cdr NIL)}
\item \texttt{(eqlablep NIL)}
\item \texttt{(index-of NIL NIL)}
\item \texttt{(index-of-aux NIL NIL NIL)}
\item \texttt{(index-of-aux-eq NIL NIL NIL)}
\item \texttt{(index-of-aux-eql NIL NIL NIL)}
\item \texttt{(len NIL)}
\item \texttt{(return-last NIL NIL NIL)}
\item \texttt{NIL}
\end{itemize}
In essence, the next step is to combine these templates with each of the
2,380 previously generated templates. However, that would lead to
a veritable explosion of resulting templates, since many of the earlier
templates have more than one placeholder \texttt{NIL}, e.g., the template
\texttt{(or (equal NIL NIL) (integerp NIL))} which has three \texttt{NIL}s,
so it would result in $13^3 = 2197$ templates by itself. In a language with
lazy evaluation, this would not be a major problem, but with ACL2 we were
concerned about space limitations if we tried to generate all the templates.

Instead, DrLA proceeds as follows. For each template, it determines how many
placeholders it has, e.g., 3. Then it computes the total number of possible
results, which in this case would be $13^3=2197$. Then it uses a counter to
go through all of the possibilities, and generates the corresponding template
just-in-time. E.g., a few of the templates generated may 
be\footnote{In the list, the numbers refer to the counter in decimal and [base13].)}
\begin{itemize}
\item[0.] [000] \texttt{(or (equal (+ NIL NIL) (+ NIL NIL)) (integerp (+ NIL NIL)))}
\item[1.] [001] \texttt{(or (equal (+ NIL NIL) (+ NIL NIL)) (integerp (append NIL NIL)))}
\item[12.] [00c] \texttt{(or (equal (+ NIL NIL) (+ NIL NIL)) (integerp NIL))}
\item[13.] [010] \texttt{(or (equal (+ NIL NIL) (append NIL NIL)) (integerp (+ NIL NIL)))}
\end{itemize}
The important point is that DrLA does not generate all of these at once. 
Rather, it generates the first, fully processes it (as described below), and
then loops to generate the next template.

A further optimization is performed at this stage, which helps to manage the
combinatorial explosion. Consider a term like
\texttt{(and (integerp NIL) (integerp NIL))}. We call these terms redundant,
and the key observation is that such a term was formed from a Boolean template,
\texttt{(and NIL NIL)} in this case, and a predicate,
\texttt{(integerp NIL)} in this case. But since \texttt{NIL} is a Boolean
template in its own right, then \texttt{(integerp NIL)} will also be generated
as a possible hypothesis---thus, we can safely ignore the more complicated
but logically equivalent \texttt{(and (integerp NIL) (integerp NIL))}. It's not
just duplicates that cause this, e.g., consider
\texttt{(and (integerp X) (acl2-numberp X))} where one subterm implies the 
other. DrLA recognizes these cases and will ignore any template with a redundant
subterm, since it knows a simpler, equivalent term will also be considered.

These are still templates, however, not full ACL2 terms. The next step is to
consider the leaves that may be placed in each individual template. These are
the free variables appearing in the original theorem, and 
possibly\footnote{By default, DrLA does not use the built-in binary predicates
\texttt{equal} and \texttt{<<}, in which case it also does not consider
built-in constants. This reflects an attempt to manage the combinatorial
explosion of term generation.} a handful of
constants such as 0, 1, \texttt{t}, \texttt{nil}. In our running example, the 
variables are \texttt{K}, \texttt{X}, and \texttt{Y}.
The final step in term generation is to use these variables and constants to
fill in the placeholders in the templates generated above. Again, this is done
lazily by filling in each template with each possible n-tuple of values before
moving to the next. The resulting terms will include the following (and many,
many more):
\begin{itemize}
\item \texttt{(or (equal (+ X X) (+ X X)) (integerp (+ X X)))}
\item \texttt{(or (equal (+ K Y) (+ Y X)) (integerp (+ X K)))}
\end{itemize}

DrLA then processes each term to determine which should be candidate
hypotheses. Recall that DrLA started by gathering a set of counterexamples
and witnesses from cgen. Each counterexample or witness consists of a 
list of bindings for the free variables in the original theorem. E.g.,
a counterexample may look like
\begin{verbatim}
    ((K 3) (X '(1 2 3)) (Y '(10 20 30)))
\end{verbatim}
and a witness may look like
\begin{verbatim}
    ((K 10) (X '(1 2 3)) (Y '(10 20 30)))
\end{verbatim}
The important point is that witnesses and counterexamples fully bind the
terms under consideration, so DrLA can use ACL2's executable interpreter to
determine the value of each term under that assignment. We use
\texttt{trans-eval} for this purpose, although there are other hooks into
the ACL2 interpreter. What remains, then, is to check if the term evaluates
to false for all of the counterexamples. If that's the case, then
\begin{verbatim}
    (implies TERM
             BROKEN-THEOREM)
\end{verbatim}
may be an actual theorem, so the term may be a good candidate hypothesis.
At the very least, it eliminates all the (known) counterexamples.

Of course, we want the term to be satisfiable so that adding it as a
hypothesis does not make the final theorem vacuously true. This is where
the witnesses come in. Ideally, the proposed hypothesis will be true for
all the witnesses. Such a hypothesis neatly separates the witnesses from
the counterexamples (always true for the former and false for the latter).

However, we found that this is not always desirable. Our intuition is that
when a user submits a theorem to ACL2, she has an expectation of the types
of the variables under consideration, or some constraint on the possible
values of them. The reason the proof attempt failed, however, is that this
expectation was not explicit in the hypothesis, so the theorem as stated is
in fact false. The counterexamples clearly attest to that, but the witnesses
do not necessarily do the same. For example, consider the (false) theorem
\begin{verbatim}
    (equal (<= (* k x) (* k y))
           (<= x y))	
\end{verbatim}
Here, the user forgot to specify that \texttt{K} is a non-negative integer.
However, cgen may find witnesses such as
\texttt{k=-1}, \texttt{x=0}, and \texttt{y=0}.
In fact, it may find witnesses such as 
\texttt{k=-1}, \texttt{x=NIL}, and \texttt{y=NIL}.
We call these witnesses, which are necessarily outside of the user's 
intent, ``false witnesses.''
Insisting that any proposed hypothesis fully separates the counterexamples
from the witnesses is too strict a requirement in the presence of
false witnesses, so DrLA considers a hypothesis to be a good candidate
if it is false for all the counterexamples, and true for at least one witness.

As mentioned previously, DrLA performs a subsumption check to eliminate as
many candidate hypotheses as possible. The subsumption check is not just
syntactic (as in traditional subsumption) but semantic, since it involves
logical entailment. It would be possible to use the theorem prover itself
to determine this, but we chose not to for two reasons. First, the subsumption
checks happen often, as each potential hypothesis is compared to all other
hypotheses. Making these many calls to the prover may simply prove
impractical. Second, the theorem prover is not complete, so it may fail to
prove that one hypothesis does in fact subsume another. Instead, we chose to
use the witnesses and counterexamples gathered from cgen to test logical
entailment using the executable interpreter in ACL2. The same tradeoff
calculation led to using the interpreter to determine when a template is
redundant, as mentioned previously. Additional experiments are necessary
to determine the right tradeoffs, so these design decisions may change in the
future.

The last step is reporting the final list of candidate hypotheses to the
user. Here, DrLA invokes the theorem prover on each ``fixed'' theorem, to see
which of the suggested hypotheses in fact succeed in fixing the proofs. DrLA
displays all suggestions to the user, not just the ones that resulted in
successful proof attempts, since the real goal isn't a ``Q.E.D.'', but
making explicit the hidden assumption that the user had forgotten. It may well
be that one of the failing suggestions is actually closer to what the user
intended, and seeing that suggestions will inspire the user to correct the
theorem.

\section{Conclusion and Further Improvements}
\label{sec:conclusion}

This paper introduced DrLA, a tool that can help ACL2 users fix broken
theorems by suggesting a forgotten hypothesis. This is done by borrowing
and repurposing ideas from theory exploration and machine learning.

Our experience suggests that DrLA is a promising tool, though more work
is needed before it can reach its potential. Some problems are technological.
For example, since DrLA relies on counterexample generation, it can be only
be used on formulas that are fully executable, i.e., no encapsulates. This
problem may be addressed effectively using defattach.

Other problems are about efficiency. There is a combinatorial explosion at
the heart of DrLA. Computers are getting faster, but combinatorics can't just
be brushed aside. There are certainly more algorithmic tricks we can use to
reduce the number of templates generated. The reader may have noticed, for
example, that the templates described in Sect.~\ref{sec:implementation} had
entries for both \texttt{+} and \texttt{binary-+}. Detecting such redundancies
could result in significant speedups.

Other improvements may come from revisiting DrLA's heuristics. This is
particularly true of the selection of function names to be used in the
term templates. DrLA looks for function symbols in the theorem itself, and in
the definitions of functions used in the theorem. Some of this is necessary,
or DrLA would only ever be able to suggest simple typing hypotheses. But as the 
examples showed, DrLA wastes a lot of time considering possible terms that
are highly improbable, e.g., \texttt{eqlablep} or \texttt{index-of-aux-eql}.

On the other hand, DrLA sometimes fails to find any suitable hypotheses simply
because it does not know to use certain functions. One possible way of 
addressing this is to follow the strategy of the tau system, which is to
consider all unary predicates. (It is worth noting that unary predicates are
especially effective at battling the combinatorial problem, since adding a
unary predicate does not change the number of available slots.) However, this
too could result in an excess of templates, since unary predicates are very
common, and DrLA can't use the greedy optimizations that are so effective in
the tau system. One possibility is to use an explicit system, as in the
\texttt{defdata} types of ACL2. That is something we plan to explore in the
near future.
Another possible solution can be found by using machine
learning, which is part of the broader context in which DrLA was developed. In
particular, machine learning could be used to explore the Community Books for
clusters of predicates in hypotheses that are associated with other predicates
in conclusions of theorems. For example, theorems about binary trees may often
use hypotheses about balanced trees. Even if ``balanced'' is not mentioned in
the conclusion, DrLA could use such information to automatically consider it
in such cases.

Finally, another place where DrLA could be improved is in the interaction with 
the counterexample generator. Cgen makes extensive use of the known datatypes
in the way that it searches for counterexamples. For instance, if the formula
contains a variable \texttt{X} that is known to be a binary tree, cgen will
try to find witnesses and counterexamples where \texttt{X} is bound to a
binary tree. But this information is made available to cgen through the
related \texttt{defdata} framework, which not all ACL2 users currently use.
Addressing this is a major challenge, but one that is worthwhile. A better
way of finding witnesses and counterexamples will immediately upgrade DrLA.

DrLA is an ongoing project, and we hope to continue improving it in the near
future by following the roadmap described in this section.

\nocite{*}
\bibliographystyle{eptcs}
\bibliography{drla}
\end{document}